\documentclass[9pt,twocolumn,twoside]{pnas-new}

\usepackage{graphicx}
\usepackage{tabularx}
\usepackage{authblk}

\newcommand{\bfr}{{\bf{r}}}
\newcommand{\epsM}{{\varepsilon_M}}

\templatetype{pnasresearcharticle}

\title{Micro-slips inside a granular shear band as nano-earthquakes}

\author[a]{David Houdoux}
\author[a]{Axelle Amon}
\author[b]{David Marsan}
\author[c]{J\'er\^ome Weiss}
\author[a]{J\'er\^ome Crassous}

\affil[a]{Univ Rennes, CNRS, IPR (Institut de Physique de Rennes) - UMR 6251, F-35000 Rennes, France}
\affil[b]{Universit\'e Savoie Mont-Blanc, CNRS, IRD, IFSTTAR, ISTerre, Le Bourget-du-Lac, France}
\affil[c]{IsTerre, CNRS/Universit\'e Grenoble Alpes, 38401 Grenoble, France}

\significancestatement{Earthquakes are caused by sudden releases of energy along faults. Although hardly predictable, their statistical properties follow robust empirical laws highlighting a complex organization of the frictional forces along the faults. Here we present an experimental analog of a fault which reproduces this full complexity at the laboratory scale. By looking at the localized displacements into a sheared disordered material, we identify elementary slip events. They organize spontaneously along faults, and their statistical properties follow the empirical laws of natural earthquakes. This complex spatio-temporal organization of elementary slip events results from cascades of event triggering as earthquakes. Our work opens the door towards a better understanding of earthquake physics from the possibility to control and monitor an experimental model fault.}
\correspondingauthor{\textsuperscript{2}To whom correspondence should be addressed. E-mail: jerome.crassous@univ-rennes1.fr}

\begin{abstract}
We study experimentally the fluctuations of deformation along a shear fault naturally emerging within a compressed frictional granular medium. Using laser interferometry, we show that the deformation inside this granular gouge occurs as a succession of localized micro-slips distributed along the fault. The associated distributions of released seismic moments, the memory effects in strain fluctuations, as well as the time correlations between successive events, follow exactly the empirical laws of natural earthquakes. Using a methodology initially developed in seismology and social science, we reveal, for the first time at the laboratory scale, the underlying causal structure. This demonstrates that the spatio-temporal correlations of the slip dynamics effectively emerge from more fundamental triggering kernels. This formal analogy between natural faults and our experimentally controllable granular shear band opens the way towards a better understanding of earthquake physics. In particular, comparing experiments performed under different imposed deformation rates, we show that strain, not time, is the right parameter controlling the memory effects in the dynamics of our fault analog. This raises the fundamental question of the relative roles of strain-dependent structural rearrangements within the fault gouge vs that of truly time-dependent, thermally activated processes, in the emergence of spatio-temporal correlations of natural seismicity.
\end{abstract}

\dates{This manuscript was compiled on \today}

\begin{document}

\maketitle
\thispagestyle{firststyle}
\ifthenelse{\boolean{shortarticle}}{\ifthenelse{\boolean{singlecolumn}}{\abscontentformatted}{\abscontent}}{}

\dropcap{E}arthquakes are natural phenomena displaying scale-free
statistics~\cite{scholz}. Empirical power laws are observed for the
distribution of their moments (Gutenberg-Richter's law), rupture lengths and durations, rupture slips ~\cite{Mai.2002}, temporal~\cite{Kagan.1991} and spatial correlations between earthquakes~\cite{Kagan.2007}, which also express through a decaying rate of aftershocks (Omori's law)~\cite{omori.1894}, characterized by a scale-free (sub)diffusion~\cite{Tajima.1985,Marsan.2000}. Understanding the origin of those laws as
well as reproducing them at the laboratory scale remain nowadays major
issues. From a fundamental point of view, these scaling laws are reminiscent of nonequilibrium dynamics and critical phenomena~\cite{sethna2001,salje2014}, and
raise the question of the existence of a universality class to which
earthquakes would belong. Among mechanical systems, possible candidates
for such a universality class are those for which deformation occurs
through avalanches. The mechanical response of those systems is
characterized by an intermittent dynamics alternating slow elastic loading
and rapid sliding and relaxation, leading to a jerky dynamics and/or stress drops.

Power-law distributions of slip sizes or relaxed energies have been evidenced experimentally in various systems, such as the stationary propagation of a fracture front in a heterogeneous material~\cite{bares2018}, compression of heterogeneous materials~\cite{baro2013,vu2019}, or systems involving frictional sliding on or within a granular media~\cite{zadeh2019,Riviere.2018}. The common basic ingredients underlying the dynamics of those different systems are the existence of material disorder and the decomposition of the
dynamics in elementary events localized both in space and time, coupled together
by elasticity. A progressive evolution of avalanche size and duration statistics has been reported for different heterogeneous materials~\cite{vu2019} or granular media~\cite{denisov2016} upon increasing the loading up to a macroscopic yield or failure stress at which scale-free statistics are observed, arguing for a 'stress-tuned' critical behavior fundamentally different from a self-organized critical dynamics characterized by steady-state statistics~\cite{uhl2015}. It has been proposed on the basis of a mean-field model of plasticity that those different systems, as well as deformed microcrystals and earthquakes, could belong to the same class of universality~\cite{uhl2015,dahmen2011}. This, however, was mainly addressed from an analysis of the size distribution of avalanches as well as their average shape. In case of earthquakes, the stress-tuned critical hypothesis was argued~\cite{uhl2015} on the basis of a dependence of magnitude distributions on the slip direction on the fault plane (the rake angle), which gives indirect information about the differential stress acting on the fault~\cite{schorlemmer2005}. However, if seismic moment distributions appear to be indeed exponentially tapered at very large scales, the associated upper corner magnitude was found to be independent of the region or the depth interval considered, or of the plate velocity, i.e. to be rather 'universal'~\cite{kagan.2002}.
On the other hand, an important and ubiquitous feature of brittle deformation in the crust is the existence of aftershocks, which occurrences are also governed by scale-free laws. Much fewer attempts have been done to model those spatio-temporal correlations, through the introduction of a memory mechanism such as a slow healing of frictional properties~\cite{benzion2011}, or viscoelastic relaxation~\cite{jagla2014}. Similarly, lab experiments reproducing the clustering of events in time and space remain scarce.

As a matter of fact, systems displaying avalanches can have also fundamental differences that limit the pertinence of a universal picture. As already noted, systems in a
stationary regime must be distinguished from those for which the
spatio-temporal dynamics is evolving (stress-tuned). In case of
frictional granular media, besides a progressive increase of the maximum avalanche size~\cite{denisov2016}, the spatial distribution of plastic events evolves during loading: initially homogeneously
distributed in the bulk of the material, plasticity progressively localizes to form shear bands at macroscopic yield~\cite{lebouil.2014,Gimbert.2013,karimi2019}, in which all the shear rate concentrates afterwards while
the rest of the system becomes an elastic 'solid'. Identifying
clustering in time is only possible within those shear bands, when the spatio-temporal
organisation becomes stationary. For stationary systems,
the dimensionality of the active zone is expected to play a role, at
least on the value of the critical exponents. One must thus
distinguish tri-dimensional systems (e.g. plasticity distributed in the
bulk of an amorphous material), from those where the plasticity is confined
to a quasi-2D zone (a fault in the case of earthquakes, a shear band
in the case of amorphous granular media),
and finally quasi uni-dimensional active zones (e.g. a propagating
crack front). Practically, in experimental works pertaining to the plasticity of
amorphous media, it is not always clear whether the plasticity is
broadly distributed in the bulk of the system or if it is localized along
a shear band. Even when the geometry of the active zone is identified,
most experimental set-ups are unable to fully resolved the
spatio-temporal organization of the avalanches which are solely
identified and studied through indirect measurements of their sizes such as acoustic
emissions(e.g.~\cite{Riviere.2018}) or stress drops on a loading curve(e.g.~\cite{denisov2016}).

To address the challenging issue of reproducing an analog of a fault gouge at the
lab scale, a straightforward approach consists in imposing the bi-dimensional geometry in stationary conditions by confining a granular material between elastic
plates~\cite{geller2015,lieou.2017,Riviere.2018,lherminier2019}. In the vast
majority of those experimental studies, quasi-periodical stick-slip events with a typical size are observed, indicating that finite size effects dominate the dynamics~\cite{degeus2019}. The slip events then involve the whole length of the shear band and the dynamics loose its universal features. In addition, those macro-slip events are characterized by a reverse asymmetry of the activity compared to earthquakes, with foreshocks of increasing size as approaching the macro-instability, but an absence of aftershocks~\cite{Riviere.2018}, likely resulting from the finite-size effects mentioned above. A recent experiment with a quasi-2D shear cell in a stationary regime displayed an intermittent dynamics sharing several features of earthquakes dynamics, such as the G-R law and a power law decay of the rate of 'aftershocks'~\cite{lherminier2019}. While giving promising results in terms of the analogy between the dynamics of stationary sheared granular materials and that of earthquakes, it did not give a direct characterization of the localization and the spatial extension of the detected events. Consequently, the 'aftershock' characterization amounts to a time correlation analysis of discrete events, without quantifying the underlying \textit{causal} triggering. We can thus ask: Is it possible to build a laboratory analog of a fault gouge where well-identified events would share all the properties of earthquakes, and more particularly their spatio-temporal, scale-free clustering properties arising from stress transfers and the resulting cascades of triggering~\cite{helmstetter2005,marsan2008,marsan2010,kagan}.

Here we present experimental results obtained in a 3D granular system in a post-(macro)yield regime displaying a stationary shear band and in which finite-size effects do not dominate the dynamics. Using an interferometric method of measurement of micro-deformations we provide direct spatially-resolved measurement of the micro-slip events that govern the frictional motion along the shear band. We are able to measure the localization, the spatial extension and the magnitude of those events, providing the first direct experimental measurement at the laboratory scale of frictional micro-slips along a fault. We show that the statistics of those events displays scale-free behavior in close agreement with earthquake phenomenology. Using a methodology developed for earthquake analysis~\cite{marsan2008}, we go another step further compared to previous experimental studies by quantifying the causal triggering between events. We show that this underlying triggering process can explain the observed space-time correlations in the dynamics, much like it does for earthquakes. We argue on this basis that a frictional shear band in a granular material represents a formal analog of tectonic faults, with an intermittent dynamics probably belonging to the same universality class.

\section*{Strain fluctuations inside a shear band}

\subsection*{Stationary shear band and strain imaging}

\begin{figure*}[htbp]
\centering
\includegraphics[width=\linewidth]{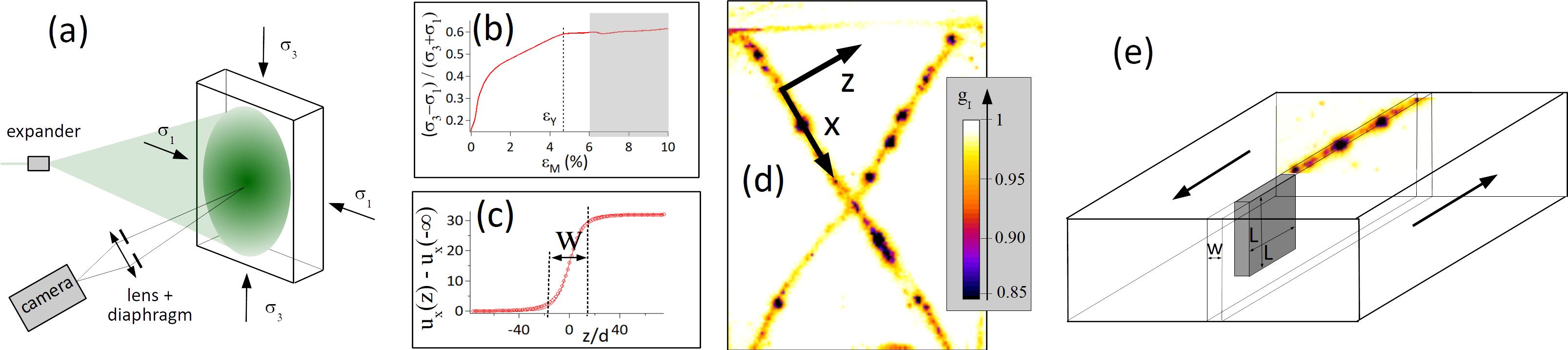}
\caption{{\bf Imaging shear band fluctuations} (a) Schematic of the
  experimental set-up. The material is submitted to a biaxial stress
  test. The front face of the sample is imaged on a camera. As
  illumination is done using coherent light, those images display
  speckles. (b) Normalized deviatoric stress as a function of macroscopic deformation. $\varepsilon_Y$ is the yield strain, and the gray zone is the post-yielding zone analyzed in this study. (c) Relative displacement $u_x$ of two blocks separated by a shear band as a function of the direction perpendicular to the shear band. Symbols are experimental data, plain line is
  $u_x(z)=\Delta u_x \times [1+\tanh(-2z/w)]/2$, with $\Delta u_x=32~nm$
  and $w=22~d$. (d) Map of the correlation between two successive speckle images.
  The color of the pixel is related to the value of the correlation.
  (e) Schematic of the shear band separating two sliding
  blocks as composed of discrete shear events of fixed width $w$ and of size $L\times L$.}
\label{fig1-2}
\end{figure*}

We use for this study an experimental set-up which consists of a
biaxial cell filled with a granular material
composed of an assembly of glass beads confined into a rectangular
box (see fig.~\ref{fig1-2}a). The two lateral faces are
deformable latex membranes which allow us to impose a confining stress $\sigma_1$ to the material. This stress is kept constant during the full experiment. The material is slowly compressed by moving at a fixed
velocity the top face with respect to the bottom face, and the axial
macroscopic deformation $\varepsilon_M$, and the applied axial stress
$\sigma_3$ are measured (S.I.1,~\cite{Lebouil.2014a}).

The strain fluctuations are imaged using an interferometric technique
based on Diffusing Wave Spectroscopy. For this, the material is
illuminated with an extended laser beam, and the speckle images are
regularly recorded. We note $\delta \varepsilon^*_M$ the macroscopic
strain increment between two successive images, and its value is $\delta \varepsilon^*_M=5. 10^{-7}$ if not otherwise specified. The speckle
images are then divided into square zones, and for two successive
images, the normalized autocorrelation function of the scattered
intensities are calculated for each zone. Associating a color to the
value of the correlation at a position, we obtain maps of correlation
$g_I(\varepsilon_M,\bfr)$, where ${\bf r}$ is the position on the observation plane, as shown on fig.~\ref{fig1-2}d. High correlation $g_I \approx 1$ (white pixels) indicates that beads are
uniformly translated without relative motions, whereas low correlation
$g_I << 1$ (dark pixels) is the signature of bead relative
motions. In addition to this interferometric correlation technique,
we use a conventional digital image correlation method on the speckle
pattern: the displacement of zones of the speckle
pattern between different images are measured, giving access to the
displacement field. This measure is used to determine the mean
relative velocity of blocks when shear bands are formed.

Starting from an initial condition of a material submitted to a
isotropic confining pressure, the material is slowly compressed. The
beginning of the compression is associated to a plastic flow spatially
distributed into the sample and to an increase of the stress difference $\sigma_3-\sigma_1$ (see fig.~\ref{fig1-2}b). We analyzed this plastic flow in previous works~\cite{lebouil.2014} (see also Movie in SI.) and we do not look further to this initial stage here. When the deformation
$\varepsilon_M$ of the material exceeds a yield strain $\varepsilon_Y\approx 4.5\%$, the stress difference $\sigma_3-\sigma_1$ is roughly constant (see fig.~\ref{fig1-2}b), and the deformation localizes into the material, with the formation of one or two linear shear bands~\cite{nguyen2016}. The shear is not localized close to a
moving mechanical boundary as it is the case for a Couette cell (where
shear band appears at rotor), or in a  gouge confined between
two rigid blocks. Here, the shear band emerges spontaneously in the
system. Its orientation is linked to intrinsic properties of the
material (it is linked to the deviatoric stress at failure through the
Mohr-Coulomb relationship) and not to geometrical constrains. Its
width is the result of the self-organization of the system: the flowing
material forming the band and the solid material surrounding it is the
same and this separation of phase emerges spontaneously in the system.

\subsection*{Average vs instantaneous strain}
The map of the correlation of the scattered intensity can be linked to
the shear motion of the sliding blocks at each side of a band. This
may easily be seen qualitatively: for this we consider a correlation
map obtained in the stationary regime (see fig.\ref{fig1-2}d), i.e. when
the stress difference is in a plateau phase, and $\varepsilon_M >
5\%$. The correlation is close to 1 into the four triangular zones
partitioned by decorrelated boundaries. This indicates that the
material is split in four rigid blocks separated by
deformed zones.

To obtain a quantitative information about the shear field inside the
band, we assume (this hypothesis will be discussed just below) that the motion of the beads around a point $\bfr$ is
mainly a shear $\gamma_{m}(\bfr,\epsM)=\partial u_x / \partial z$, where
$\bf u$ is the local displacement of a block with respect to another
one, and $(x,z)$ are local coordinates associated to a band (see
fig.\ref{fig1-2}d for axis definition). By making this assumption, we neglect other
components of the strain tensor and uncorrelated motion of the
beads. If the beads move accordingly to this shear field, the
decorrelation can then be related to the local shear as (see SM.1.4):

\begin{equation}
\gamma_{m}(\bfr,\epsM)= - \gamma_0 \ln[g_I(\bfr,\epsM)] \label{eq:loc_strain}
\end{equation}
with $\gamma_0=2.6\times 10^{-4}$ a constant given by the optical properties of the
material.The time-average local deformation is defined as:
\begin{equation}
\bar{\gamma}_{m}(\bfr,\epsM)= (1/\Delta \epsM)\int_{\epsM}^{\epsM
  +\Delta \epsM} \gamma_{m}(\bfr,\epsM')d\epsM' \label{eq:mean_strain}
\end{equation}

The hypothesis of local shear can be quantitatively tested. For this, we
integrate~\eqref{eq:mean_strain} along a direction perpendicular to
the shear band, and we obtain
$u_x(z)-u_x(-\infty)=\int_{-\infty}^z \bar{\gamma}_m(z',\epsM)
dz'$. Figure~\ref{fig1-2}c shows the displacement field across the shear
band, demonstrating that the strain is concentrated into a narrow
zone of width $w=22d$. For our macroscopic strain increment
$\delta \varepsilon^*_M=5. 10^{-7}$, the relative sliding of the two
blocks $\Delta u_x=u_x(\infty)-u_x(-\infty)$ if found to be
$\Delta u_x=32~nm$. This value is close to the imposed value
$\Delta u_x=25~nm$ that we may estimate from the displacement of the
top plate and the orientation of the shear band. This agreement
confirms the hypothesis that the decorrelation of speckle pattern is
mainly due to the shear motion of the beads. The difference presumably
arise from some uncorrelated motion of the beads occurring in the
sheared zones.

\subsection*{Memory effect in strain fluctuation}

\begin{figure*}[htbp]
\centering
\includegraphics[width=\textwidth]{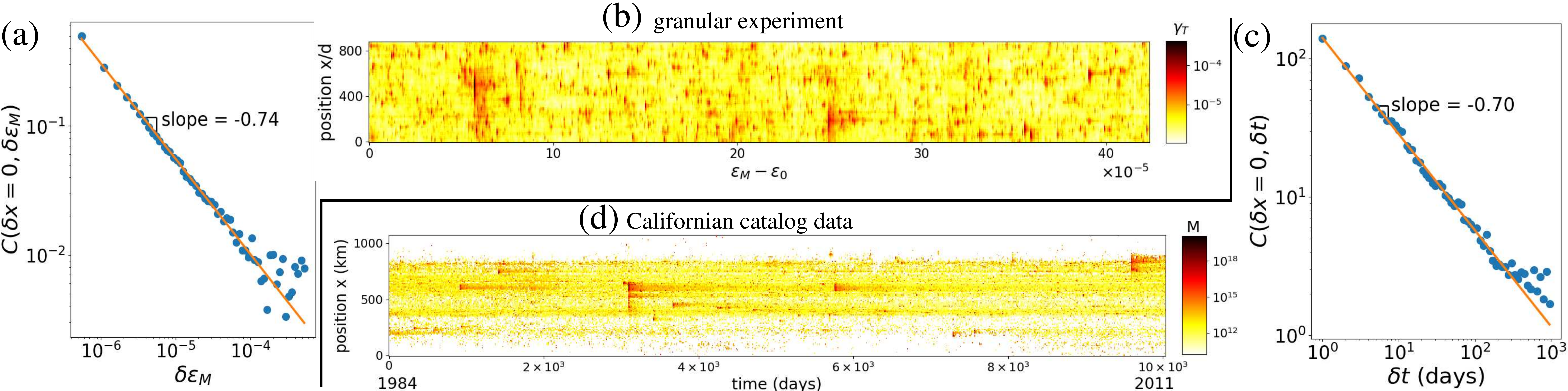}
\caption{(a) and (b): granular experiment. (a) Spatiotemporal correlation function $C(\delta x=0,\delta \epsM)$ as a function of $\delta\epsM$ see Eq.~\eqref{definitionC}. (b) Spatiotemporal evolution of the local strain $\gamma_T(x,\epsM)$ (see text) in a shear band. (c) and (d): Californian earthquake catalog. (c) Spatiotemporal correlation function $C(\delta x=0,\delta t)$ as a function of $\delta t$ (see S.I.2 for details). (d) Spatiotemporal evolution of the seismic moment $M$ resulting from earthquakes from the Californian catalog projected along the main direction of the fault (see S.I.3 for details).}
\label{fig_ST}
\end{figure*}

We now analyze the fluctuations of the local shear. Since deformation is located into the shear band, we consider the transverse-averaged shear deformation $\gamma_{T}(x,\epsM)=\frac{1}{2w}\int_{-w}^{w}\gamma_{m}(\bfr,\epsM) dz$. Figure~\ref{fig_ST}b shows $\gamma_{T}(x,\epsM)$ into the $(x,\epsM)$ plane. We can clearly see that the shear is heterogeneous both in space (i.e. along the shear band) and in time (i.e. along the macroscopic deformation). In
order to analyse those fluctuations, we introduce the normalized spatiotemporal correlation function:

\begin{equation}
C(\delta x,\delta\epsM)=
\frac{
\langle \gamma_{T}(x,\epsM)\times \gamma_{T}(x+\delta
x,\epsM+\delta\epsM) \rangle}
{\langle \gamma_{T}(x,\epsM) \rangle \langle \gamma_{T}(x+\delta
  x,\epsM+\delta\epsM) \rangle} - 1~\label{definitionC}
\end{equation}

where $\langle \cdot \rangle$ is an average both on deformation and position (see S.I.2 for details). Figure~\ref{fig_ST}a is a plot of $C(\delta x=0,\delta \epsM)$ as a function of the $\delta\epsM$. There is clearly memory in the deformation. Moreover, this correlation function decays as a power law
$C(\delta x=0,\delta\epsM)\sim (\delta\epsM)^{-\gamma}$, with $\gamma=0.74$. Binning in space and time the moment along the San Andreas Fault system from the Californian earthquake catalog (see S.I.3 for details), very similar spatiotemporal patterns (Fig.~\ref{fig_ST}d) and correlations (Fig.~\ref{fig_ST}d) are obtained. The analogy between our shear band and tectonic faults in terms of scaling laws of seismic moments, temporal clustering, and aftershock triggering, is thoroughly analyzed in what follows.

\section*{Shear band as a model fault}

\subsection*{Definition of shear transformation  events}

As we saw in~\ref{fig_ST}a, the local shear $\gamma_{T}(x,\epsM)$ exhibits important fluctuations, alternating activity and quiescent phases. The concept of shear transformation zone has been introduced to deal with the flow of disordered materials: spatial zones reorganize, creating a local shear. We define such zones from the light scattering data. For this, we apply a threshold to the image $\gamma_{m}(\bfr,\epsM)$, and we use a particle detection algorithm to obtain individual shear events (see SI.2 for details about threshold and detection algorithm). Events are numbered, and to each event $i$ are associated a macroscopic deformation $\epsM_{;i}$ at which the event occurs, a position $\bfr_i$ (defined as the barycenter of $\gamma_{m}$), a surface $\Sigma_i$ on the image, and a mean microscopic shear $\gamma_i$. We also define a 'seismic moment' as $M_i=\mu u_i L_i^2$ with $\mu$ the shear modulus of the granular assembly, $u_i$ the shear displacement, and $L_i^2$ the shear surface (see fig.\ref{fig1-2}e). For a shear zone of width $w$, we have $\Sigma_i=w  L_i$ and $u_i=w\gamma_i$, and thus $M_i=\mu \gamma_i \Sigma_i^2 /w$. The shear modulus of the granular assembly may estimated from mean field theory of granular elasticity~(eq.(14) of \cite{Makse.2004}) and is $\mu=200~MPa$ for a pressure of $30~kPa$. The shear band width $w=22~d$ is measured from the mean strain (Fig. 1c). The moment magnitude is defined as $m_{w;i}=\frac{2}{3}\log_{10}(M_i)-6.07$, with $M_i$ expressed in $N.m$~\cite{kanamori.2004}. The energy dissipated during an event is $E_i=\tau u_i L_i^2$, where $\tau$ is the shear stress along the shear band, and is then directly linked to the moment: $M_i/\mu=E_i/\tau$. The value of $\tau=33~kPa$ may be obtained from the principal stresses $\sigma_1$ and $\sigma_3$ at failure using Mohr-Coulomb construction.

\subsection*{Scaling laws of events}

\begin{figure}[ht]
\centering
\includegraphics[width=\linewidth]{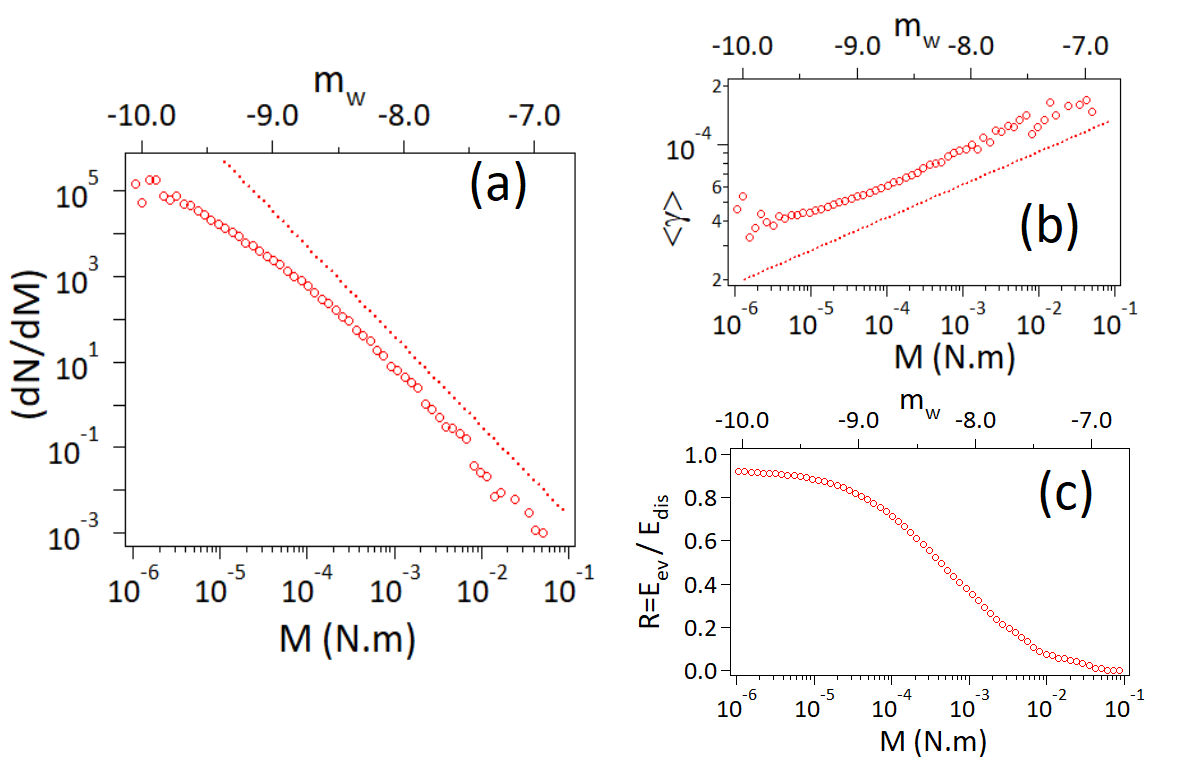}
\caption{(a) Probability density $dN/dM$ of events of moment $M$. Dotted line is a power-law $\sim M^{-2.1}$. (b) Mean deformation $<\gamma>$ as a function of the moment. Dotted line is a power-law $\sim M^{0.17}$. (c) Ratio between the energy dissipated in events of moment $\ge M$ and the total dissipated energy.}\label{fig4}
\end{figure}

We now look at the statistical characterisation of the events. We consider for this the sequence of events occurring on one half-shear band for macroscopic deformation $6\% \le \epsM \le 10\%$ (gray zone on fig.\ref{fig1-2}b). The total number of counted events is $N_{tot}\approx 1.1\times 10^{5}$. The minimum moment is dependent on the threshold and is $M_{min}=6 \times 10^{-7}N.m$, whereas the largest events have $M_{max}=0.05~N.m$. The probability density function of energy $dN/dM$ is plotted on \ref{fig4}a, and decays as a power of energy $dN/dM \sim M^{-\beta}$, with $\beta \simeq 2.1$. Although our moments stand in a range roughly 20 orders of magnitude below that of earthquakes, this behavior is similar to the empirical Gutenberg-Richter's law. Indeed, the number $N(M)$ of earthquakes with a moment magnitude larger than $m_w$ is $\log_{10}(N(m_w))=a-b m_w$, leading to $dN/dM \sim M^{-(1+(2/3)b)}$. The value of $b$ for faults is usually $\simeq 1.0$~\cite{scholz}, leading to a slope $\beta \simeq 1.66$. The mean deformation of an event of moment $M$ is defined as $<\gamma>=\sum_{M+dM>M_i>M} \gamma_i / \sum_{M+dM>M_i>M}$, for a small $dM$. As shown on \ref{fig4}b, this quantity is relatively constant $<\gamma>\sim M^{0.17}$. The broad distribution of the values of the moments $M_i$ (or equivalently of the relaxed energies $E_i$) is then mainly due to a broad distribution of sizes $L_i$, but not of deformation $\gamma_i$. In other words, the stress drop in a event $\Delta \tau_i=\mu \gamma_i$ has always the same order of magnitude. This is consistent with what is generally considered for earthquakes. Indeed, compilations of earthquake data argue for a scaling $M\sim L^3$~\cite{Wells.1994}, while $M=\mu L^2u=\mu \gamma L^3=\Delta \tau L ^3$, hence implying a constant stress drop. In our experiment, $\Delta \tau_i/\mu =\gamma_i \approx 10^{-4}$, this ratio being relatively close to the one typically obtained for earthquakes where $\Delta \tau/\mu \approx 3\times 10^{-5}$~\cite{dearcangelis2016}. We should however mention that, in our case, the increase of $<\gamma>$, i.e. of $\Delta \tau$, with the seismic moment, though weak, is significant (fig.\ref{fig4}b).  In case of earthquakes, a potential similar scaling would be indiscernible, owing to the large uncertainty on the estimation of the average slip and the variety of geophysical contexts. We may also consider the relative fraction of the shear stress which is relaxed during an event:  $\Delta \tau_i / \tau =\mu \gamma_i /\tau$. For large events, $<\gamma> \approx 2.10^{-4}$, whereas $<\gamma> \approx 3.10^{-5}$ for small events, hence showing that the events relax typically $\sim 0.1-1$ of the mean stress.

We finally look at the ratio of energy dissipated by events of moment greater than $M$: $E_{ev}(M)=\sum_{M_i\ge M} E_i$, compared to the total dissipated energy $E_{dis}$ (see S.I.6 for the details). The fig.\ref{fig4}c shows that typically half of the energy is dissipated in events of moment $M>10^{-4}N.m$, while all detected events account for more than $90\%$ of $E_{dis}$.This argues for a strong seismic coupling of our model fault. For tectonic faults, the coupling can vary considerably with the geophysical context, but is generally strong for interplate continental faults.

\subsection*{Temporal organization of events}

\begin{figure}[!ht]
\includegraphics[width=\columnwidth]{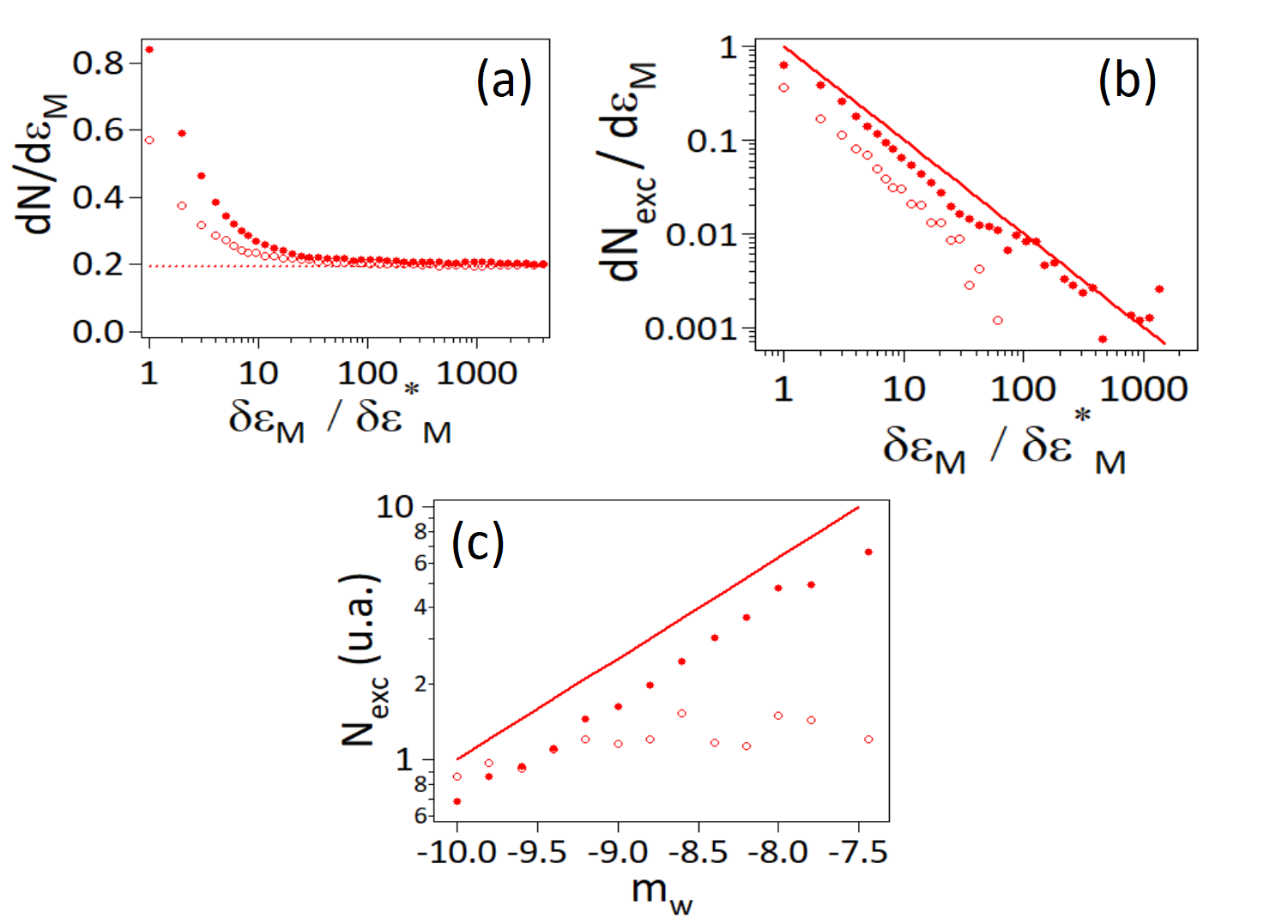}
\caption{(a) Rate of events $dN/d\varepsilon_M$ occurring after ($\bullet$) and before ($\circ$) an event of magnitude $m_w \ge -9$ as a function of the deformation increment $\delta \varepsilon_M$. Dotted line is the background rate. (b) Rate of aftershock ($\bullet$) or foreshock ($\circ$) events in excess to the background level as a function of $\delta \varepsilon_M$. Line is a $\delta \varepsilon_M^{-1}$ decay. (c) Number of events occurring in excess to background activity after ($\bullet$) or before ($\circ$) a main-shock event of magnitude $m_w$. Line is a power law $N_{exc}\sim 10^{0.4 m_w}$.}
\label{fig5}
\end{figure}

The statistical laws governing the succession of shear transformations may be analyzed within the framework of the statistical laws of natural earthquakes.

We first look at the rate of events occurring at the same position than a particular event (so-called mainshock). The figure \ref{fig5}a shows the rate of events occurring after ('aftershocks') of before ('foreshocks') 'mainshocks' of magnitude  $m_w \ge -9$ (total number of mainshocks $\approx 13 \times 10^3$). Only aftershocks or foreshock events occurring at the same position ($\pm 15d$) are counted. For large delays, the rate of events is constant, corresponding to a background $(dN/d\varepsilon)_{bg}$ rate of events uncorrelated to the mainshocks, while, at small delays, the rate excesses the background rate. The excess rate of aftershocks $(dN/d\varepsilon)_{exc}=(dN/d\varepsilon)-(dN/d\varepsilon)_{bg}$ decays with the macroscopic deformation as $(dN/d\varepsilon)_{exc}\sim \delta \varepsilon_M^{-1}$ (see \ref{fig5}b). This behavior is reminiscent of the Omori's law~\cite{omori.1894} which states that the rate of seismic events occurring after a mainshock decays with the time $t$ to the mainshock as $n(t)\sim t^{-1}$. Note however that such analysis amounts to blindly characterize time correlations between events. In particular, unlike what is generally done in earthquake analysis, the magnitude of the "mainshock" is not prescribed to be larger than that of its "aftershocks". Such correlation analysis is a signature, but not a formal quantification, of causal triggering, which is explored in details later. An opposite evolution, reminiscent of an inverse Omori's law, characterizes, in average, an increasing rate of foreshocks before mainshocks, though with a smaller rate (\ref{fig5}a) that expresses an asymmetry of time clustering.

The productivity law describes the number of excess events in response to an event of magnitude $m_w$. For this, we integrate the total number of aftershocks in excess to the background: $N_{exc}=\int_{\delta \epsM^*}^{\infty}(dN/d\varepsilon)_{exc} d(\delta \epsM)$. Figure \ref{fig5}c shows the evolution of $N_{exc}$ with the mainshock magnitude $m_w$, and we find that $N_{exc}\propto 10^{\alpha.m_w}$, with $\alpha \simeq 0.44$. This result may be compared with the productivity law for natural earthquakes, where the number of aftershocks $n_{AS}\propto 10^{\alpha.m_w}$, with $\alpha$ in the range 0.6-1.2~\cite{Hainzl.2008}. In striking contrast, the number of foreshocks appears independent of the mainshock magnitude  (\ref{fig5}c). This is in full agreement with a previous analysis of seismic foreshocks showing that such precursory activity before \textit{any} event is a mere statistical consequence of cascades of triggering~\cite{Helmstetter.2003}. Triggering of deformation events in our system is thoroughly analyzed below.

\begin{figure}[t]
\includegraphics[width=\linewidth]{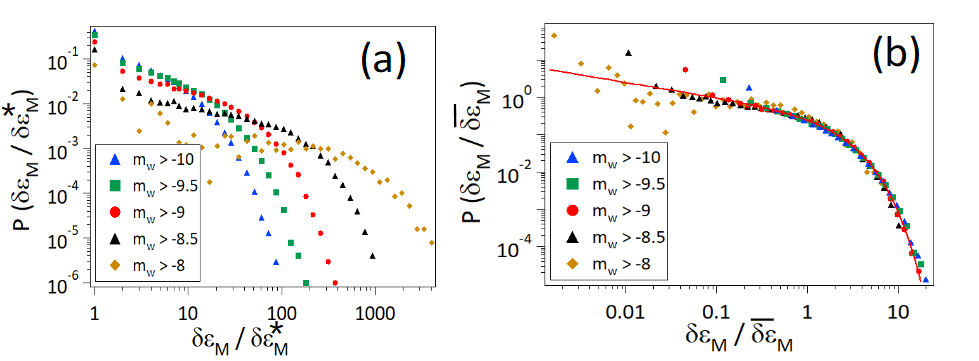}
\caption{(a) Distribution of inter-occurrence deformation between successive events, for different magnitudes. (b) Distribution of re-scaled inter-occurrence deformation $\delta \varepsilon_M / \overline{\delta \varepsilon_M}$. Plain line is a the gamma distribution $P(x)\propto x^{q-1} \exp(-x/B)$ with $q=0.6$ and $B=1.9$.}
\label{fig_intertime}
\end{figure}

The distribution of events during the loading may be further characterized by considered the first-return deformation probability $P(\delta \varepsilon_M)$ which is the analogous of the first time-return time probability for earthquakes. For this, we measure the macroscopic deformation $\delta \varepsilon_M$ between successive events, occurring at the same position ($\pm 15d$) along the band. Only events of magnitude greater than $m_w$ are considered. The figure~\ref{fig_intertime}a shows the distribution of inter-occurrence deformations for different moment thresholds. The distributions decay with a power-law, followed by an exponential decay. As shown on~\ref{fig_intertime}b, those distributions may be properly collapsed by considering for every magnitude $m_w$, the normalized deformation of $x=\delta \varepsilon_M/\overline{\delta \varepsilon_M}$, where $\overline{\delta \varepsilon_M}$ is the mean deformation between successive events. As indicated on fig.\ref{fig_intertime}b, the distribution may be well approximated by a Gamma distribution
\begin{equation}
P(x)\propto x^{q-1}\exp(-x/B)
\label{eqGamma}
\end{equation}
with $q=0.6$ and $B=1.9$. $P(x)$ decays as a power law with exponent $\approx 0.4$ up to values of $x\approx 1$, then exponential decays takes place. This behavior is not surprising, as it has been shown to be a mere consequence of a triggering dynamics characterized by a GR distribution, the Omori's and productivity laws~\cite{Saichev.2006}. Our results are very similar to observations for tectonic seismicity where $q \simeq 0.67$ ~\cite{Corral.2004}, or micro-seismicity $q \simeq 0.74$ ~\cite{Davidsen.2013}. Such scaling laws are also observed in fracture experiments~\cite{Ribiero.2015}.

\section*{Discussion}
\subsection*{Shear band viewed as a minimal model of gouge}

Starting from an initially homogeneous assembly of beads, our system organizes spontaneously to reach a stationary regime where all the deformation is concentrated along shear planes. The analysis of the statistical properties of the strain fluctuations along those planes show a strong quantitative and qualitative analogy with the statistical characteristic of natural earthquakes along a fault: the shear band may be viewed as a simplified gouge. We discuss here why this scale-free organization of deformation along a gouge is not observed in other laboratory systems.

Many mechanical systems other than crustal faults exhibit crackling noise when plastically deformed~\cite{denisov2016,baro2013,Ribiero.2015,Zadeh.2019,bares2018,Miguel.2001,lherminier2019}. In particular, the statistical properties of deformation or mechanical stress fluctuations follow power laws which reveal the absence of any particular scale in the system at least in an extended inertial (scale) range. Those fluctuations arise from individual deformations which interact elastically to create a complex scale-free dynamics. In the case of plasticity of amorphous materials before the macroscopic yielding~\cite{denisov2016,Lin.2015,uhl2015,Miguel.2001}, the criticality of the system is related to the approach of the yield point. In this case, the plastic events are expected to be initially randomly distributed throughout the bulk of the sample and not  localized in structures analog to natural faults. The brittle failure of an amorphous material is another configuration where crackling noise is observed~\cite{bares2018}. In this case, the plasticity occurs in a damaged zone close to the propagating crack tip, and a stationary plastic deformation cannot be defined. In some experiments such as compression of disordered materials~\cite{Ribiero.2015} or some granular experiments~\cite{zadeh2019,lherminier2019}, the spatial extension of events and their localization are unknown.

In order to obtain a zone of intense plasticity in a stationary regime, one can shear an artificial gouge, made of a granular material confined between elastic plates of very different elastic modulus (much stiffer or much softer)~\cite{geller2015,lieou.2017,Riviere.2018}. At first glance, this appears to be a reasonable realization of a natural gouge which consist in highly crushed rocks confined between rigid material. However, in those cases, avalanches of various sizes are not observed, but instead the dynamics is dominated by macro-slips involving all the sliding interface. De Geus~{\it et al.}~\cite{degeus2019} have recently proposed a numerical model of frictional contact consisting in an amorphous layer confined between two elastic blocks, in which scale-free dynamics and large macro-slips events implying the whole interface coexists. Such a competition between an avalanche regime and a periodic stick-slip is reminiscent of the Parkfield segment of the San Andreas Fault, confined between a creeping zone and an unloaded segment, where large and pseudo-periodic earthquakes have been observed~\cite{bakun1985}. In our experiment, the dynamics of the model gouge is dominated by scale-free avalanches and we do not observe such macro-slips. This difference of behavior may arise from the difference in confinement of the gouge. When the gouge is confined between elastic plates, there is an important contrast of mechanical properties between the gouge (which is an elasto-plastic granular material), and the plates (which are perfectly elastic). We may then expect that the plates transmit integrally the mechanical stress over all the gouge, leading to macroscopic slip events. In our experiments, materials that compose the fault and the surrounding medium are the same: both consist of the same glass beads. Given the applied pressure in our experiment, we do not expect any bead crushing, and this is in agreement with optical observations. So, the mechanical properties of the material are probably very close inside and outside the shear band, and they are both elasto-plastic. So, the material outside the gouge does not behave as a rigid block transmitting the mechanical stress on the all the interface. This is probably why we do not observe any large macro-slips but only localized avalanches following scale-free dynamics.

\begin{table*}[ht]
\centering
\begin{tabular}{|l|c|c|}
\hline
Property & earthquakes & our experiment \\
\hline
Temporal correlation function: $C(\delta x=0,\delta t) \sim \delta t^{-\gamma}$ & $\gamma \simeq 0.70$ & $(^*)~~\gamma \simeq 0.74$\\
Moment or energy distribution: $dN/dM \sim M^{-\beta}$ & $\beta \simeq 1.7 $ & $\beta \simeq 2.1$ \\
Aftershocks rate: $dN_{as}/dt \sim t^{-p}$ & $p \simeq 1.0 $ & $ (^*)~~p \simeq 1.0~(^a);~1.7~(^b)$ \\
Productivity law: $N_{as} \sim 10^{\alpha m_w}$ & $\alpha \simeq 0.8 $ & $(^*)~~\alpha \simeq 0.4~(^a);~0.24~(^b)$ \\
Recurrence time distribution: $P(x)\sim x^{q-1}\exp(-x/B)$ & $q \simeq 0.7$ & $q \simeq 0.6$\\
Stress drop / shear modulus: $\Delta \tau/\mu$ & $\simeq 3.10^{-5}$ & $\simeq 10^{-4}$ \\
Branching ratio & $0.8-1.$~\cite{nandan2017} & $\lesssim 1.$\\
\hline
\end{tabular}

\caption{\textmd{Quantitative comparison between natural gouges and granular shear band. $(^*)$: with the substitution  $t \rightarrow \epsM$. $(^a)$: count of events; $(^b)$: triggering kernel.}}\label{table1}
\end{table*}

In summary, since the stationary shear band emerges from a bulk material, we are able to observe a scale free stationary dynamics occurring in a confined space. The shear band of granular material has the right dimensionality (2D shear plane in 3D space) and the right mechanical properties to accurately model a complex gouge at the laboratory scale. As a consequence, we directly observe shear events distributed along the shear plane. The statistical properties of those events are summarized in table~\ref{table1}, and their size distributions, temporal and spatial organizations, as well as correlations of displacement, are very similar to the ones observed for natural earthquakes. We demonstrate below that the analogy can be pursued one step further through a thorough analysis of triggering.

\subsection*{Triggering of deformation events}

\begin{figure}[!ht]
\centering
\includegraphics[width=0.7\linewidth]{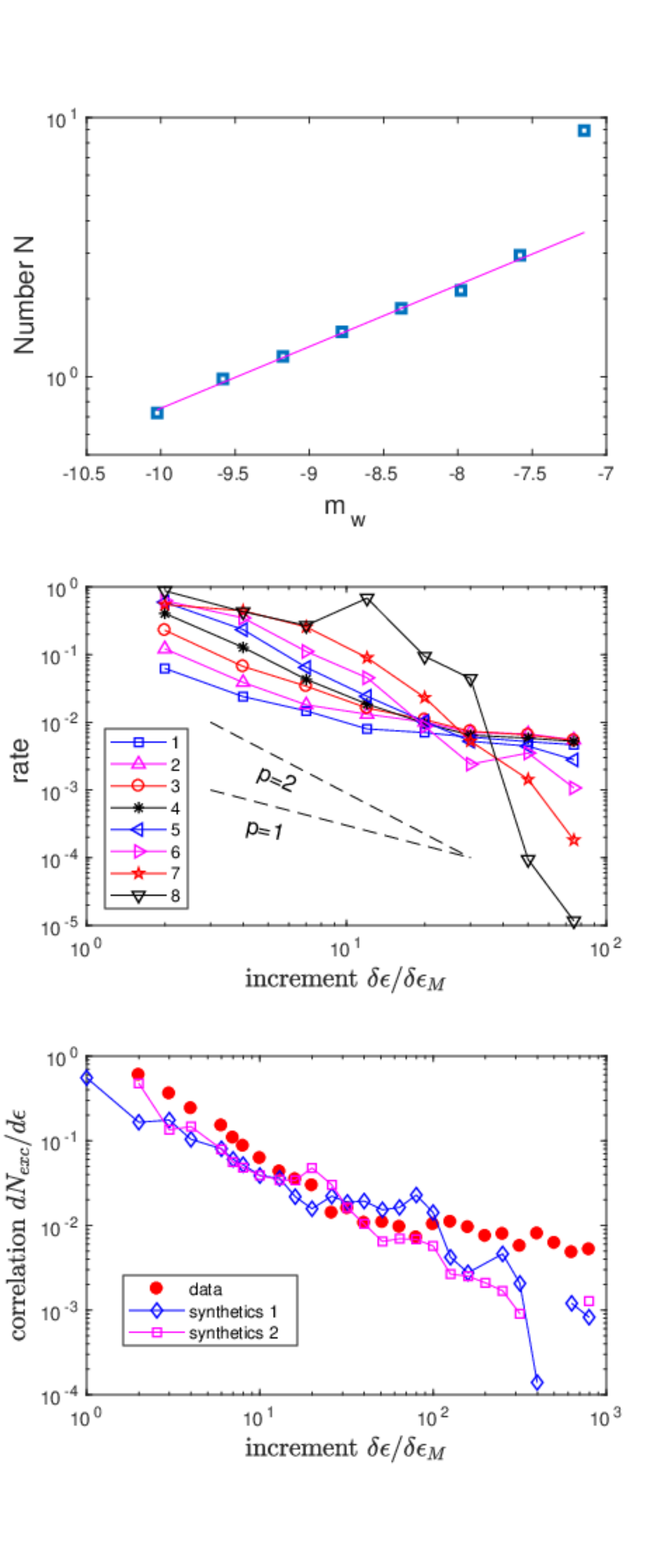}
\caption{(Top) Productivity law giving the mean number $N$ of triggered events for
a trigger of magnitude $m_w$. The best power law fit in $N \sim 10^{0.24 m_w}$ obtained when discarding the last point (biggest events) is shown in magenta. (Center) Triggering kernels in time, conditioned on the moment of the trigger. We consider the same 8 'classes' of seismic moments as in (a). Power law decays in $t^{-p}$, with $p=1$
and $p=2$, are shown for visual guidance. (Bottom) Correlation in time, as in Figure \ref{fig5}b, for two instances of synthetic datasets and for the real dataset.}
\label{fig david}
\end{figure}

Correlations in the deformation field and among deformation discrete events
are found both in time and in space, and obey power law regimes that
highlight the scale invariance of the system. However, correlation is distinct
from causality, which in the present context is equivalent to triggering, i.e.,
how the occurrence of a deformation event mechanically triggers subsequent
deformation events. The underlying causal structure can be inferred from the
data using methods that have been developed in seismology~\cite{marsan2008,marsan2010,lengline2017} or in social science~\cite{mohler2011}. We find that triggering obeys a scale-free productivity law, so that the number $N$ of directly triggered events, per mainshock, depends on magnitude as $N \sim 10^{0.24 m_w}$ (Figure \ref{fig david}, top graph), along with an Omori-like kernel, albeit with a relatively steep decay, the density of triggering events decreasing with time $t$ after the trigger as $t^{-p}$ with $p$ in the 1.6 to 1.8 interval, cf. Fig.\ref{fig david}, middle graph. Departure from these power laws is
observed for the biggest events, that produce relatively more 'aftershocks' in the early
times, but are then followed by a clear activity shutdown, both features being
likely due to a finite size effect and an exhaustion of the stressed, ready-to-fail
patches along the deformation band after such large events.
It is customary in the framework of these models to define a so-called branching ratio, which measures the capacity of a perturbation to sustain itself over potentially an infinite time (if the branching ratio is close to 1) or instead to die off quickly (if it is close to 0). This ratio can be estimated as the number of directly triggered aftershocks per 'mainshock' averaged over all the events of the catalog (REFS). We here find that the branching ratio is very close to 1, implying that the background (i.e., non-triggered) activity consists of a few $\%$ at most, so that most of the activity is made of events triggered by preceding events, highlighting the dominance of triggering and therefore clustering in the dynamics of deformation events. This is also fully consistent with earthquakes, for which the branching ratio ranges between 0.8 and 1~\cite{nandan2017}.

The productivity in $10^{0.24 m_w}$ found here is distinct from the $10^{0.40 m_w}$ scaling observed when stacking the activity past all $m_w \geq 9$ events as in Figure \ref{fig5}c; this is due to the fact that, in the latter case, the stacking mixes causally-triggered sequences (e.g., if A triggers B that triggers C, then both B and C will show up in the counting of subsequent activity, while in Figure \ref{fig david} B is counted for A while C is counted for B). We thus checked that this mixing does indeed re-create the observations of Figure \ref{fig5}. To do so, we exploit the fact that the causal structure can be formulated as a
linear model, that is simply amenable to simulations~\cite{marsan2008,marsan2010,lengline2017}. We thus simulate synthetic datasets of deformation events based on this model and its basic ingredients: (1) seismic moments are independent, identically distributed, and follow the Gutenberg-Richter-like marginal
distribution of the experiment; (2) a small proportion (about $5\%$) of the events
occur randomly in space and time, and correspond to 'spontaneous' events, i.e., events
that are not triggered by previous events; (3) the $95\%$ other events are
triggered events from previous 'mainshocks', their distribution relative to the
time and position of the mainshock following the kernels observed for
the experiment dataset (e.g., the temporal kernel of fig.\ref{fig david}, middle graph).
Generating such synthetic datasets, we find that the correlations (i.e., stacked rates) seen for
the real data are indeed well recovered (they are within the natural fluctuations of simulation outcomes), demonstrating that these correlations effectively emerge
from more fundamental triggering kernels, cf. fig.\ref{fig david} (bottom graph).

\subsection*{Structural vs temporal memory effects}

\begin{figure}[ht]
\centering
\includegraphics[width=0.9\linewidth]{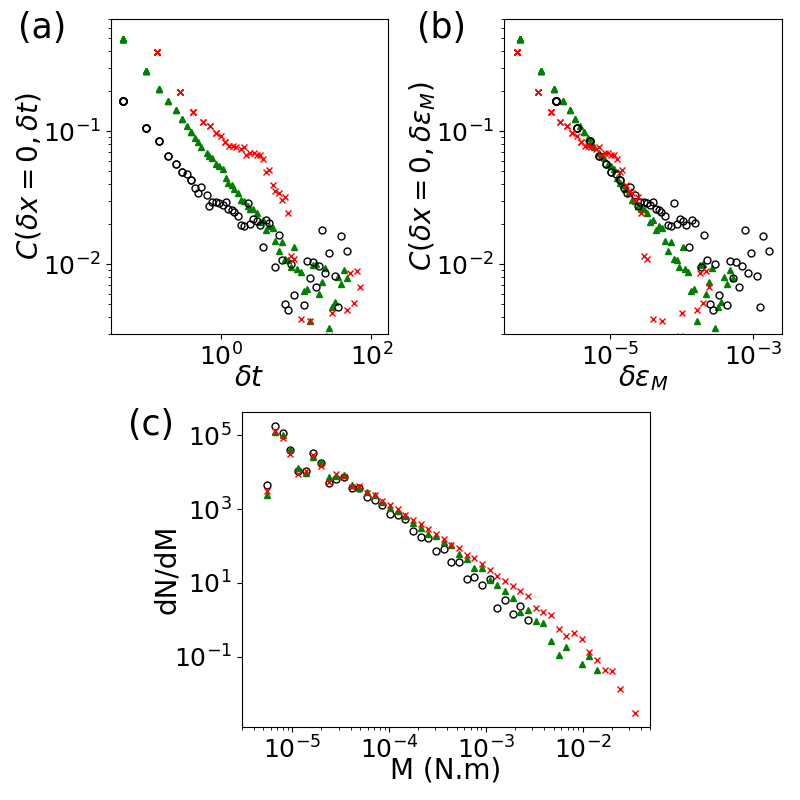}
\caption{Comparison of three experiments performed at different compression velocities: red crosses $\dot{\varepsilon}_M = 3.5\times 10^{-6}$s$^{-1}$; green filled triangles $\dot{\varepsilon}_M = 1.1\times 10^{-5}$s$^{-1}$; open black circles: $\dot{\varepsilon}_M = 3.5\times 10^{-5}$s$^{-1}$. (a) Normalized correlation function of the deformation~\eqref{definitionC} as a function of the time increment $\delta t$. (b) Same data plotted as a function of the strain increment $\delta \varepsilon_M$. (c) Probability density $dN/dM$ of events of moment $M$. (strain increment $\delta \varepsilon_M^*=1.5\times 10^{-6}$, event threshold (see S.I.5) $\gamma_s=\gamma_0$)}
\label{temporal}
\end{figure}

Strain correlation functions (Fig.~\ref{fig_ST}), rate excess after main-shocks (Fig.~\ref{fig5}a), and the causal triggering kernel (Fig.~\ref{fig david}) all indicate the existence of scale-free memory effects in our system. In the context of seismology, memory effects are remarkably revealed by the existence of aftershock sequences which are quantified by Omori's law stating that the number of aftershocks decays as the inverse of the time elapsed since the mainshock~\cite{scholz}. This law assumes implicitly that the time is the physical variable that governs the memory. The origin of such temporal dependence is however unclear. Several mechanisms such as the temporal dependence of microscopic friction law ~\cite{dieterich1972,scholz1998,gomberg2000}, sub-critical crack growth, the occurrence of afterslip~\cite{Perfettini.2007}, or poro-elasticity and the evolution with time of pore fluid pressure~\cite{Nur1972} have been proposed as a possible sources of temporal memory effects controlling earthquake occurrences. However, the direct links between any time-dependent microscopic mechanism and memory effects in seismicity are still debated.

In our experiment, we can test whether time is indeed the right parameter to describe the memory effect that we observe. For this, we performed experiments at different macroscopic deformation rates.  Figure~\ref{temporal}a shows the normalized correlation functions of the microscopic deformation expressed as a function of the time increment. If every experiment shows a memory effect, the magnitude of memory depends on the strain velocity. At a fixed time delay $\delta t$, the correlation function decreases with the velocity. This reveals that time does not seem to be the right parameter to describe memory. This may be evidenced by plotting the correlation functions as a function of the macroscopic strain increment $\delta \varepsilon_M$~Fig.~\ref{temporal}b. In this plot, the curves collapse, demonstrating that the correlation function decays with the strain increment and not with the time increment. This independence on the shear rate may also be evidenced by considering the similarity of the size distributions of events on fig.\ref{temporal}c.

This laboratory observation is evidently not in contradiction with Omori's law. Indeed, the driving velocity of a given fault is constant on the temporal scale of human observations. So, describing memory effects in terms of time increment or in terms of strain increment are then equivalent for natural faults. The fact that the memory is strain-dependent rather than time-dependent in our system suggests that the memory could be linked to structural/topological rearrangements within the granular medium, inducing a redistribution of local stress and possibly triggering slip events. It is then not surprising that the macroscopic deformation may be the parameter that governs the plasticity around a given position in the material. This also raises the question of the potential role of such geometrical rearrangements in the "time" correlations characterizing natural seismicity. In that case, such mechanisms could combine with truly time-dependent, thermally activated processes such as sub-critical crack growth, to explain memory effects in earthquake occurrences. Interestingly, slip-dependent and time-dependent memory effects combine as well in the classical Rate-and-State friction laws~\cite{Dietrich.1979,Baumberger.2006} that remains nowadays a classical framework of interpretation of earthquake physics~\cite{scholz1998,Helmstetter.2009}.

\section*{Conclusion} By looking at the intermittent strain fluctuations, we showed that a shear band inside an athermal disordered material is an analog of a natural fault: the deformation consists of many micro-slip occurring along a plane, and their collective dynamics is characterized by statistical properties remarkably consistent with the empirical laws of seismology. This analogy with natural faults is obtained when the fluctuations are observed after macroscopic yielding of the granular medium, when a steady-state regime takes place. The laboratory and natural fluctuations observed are then characteristic of a critical behavior after yielding, which is presumably different from the stress-tuned criticality observed for many systems before yielding. Our statistical analysis of micro-slips also quantifies the causal triggering between events, and reveals that this underlying triggering mechanism is at the root of the space-time correlations in the dynamics, as it has been previously shown for natural earthquakes.

Despite the simplicity of our experimental model, the phenomenological laws of seismology emerge spontaneously. This important simplification suggests that those laws may be reproduced by using simple numerical or theoretical systems of frictional particles. The analysis of such systems should allow to understand the organization of the  microscopic stress field close to the shear band. Experimentally, the possibility to control a model fault in the laboratory also opens the road to many studies, such as the effect of mechanical noise on the size distribution of slip events, the influence of the elastic properties of the surrounding material with respect to that of the band, or a study of size effects.

\acknow{The authors thank David Amitrano, Ana\"el Lema\^\i tre, Damien Vandembroucq for scientific discussions, and Patrick Chales for technical support. We acknowledge the funding from
Agence Nationale de la Recherche Grant No. ANR-16-CE30-
0022.}

\showacknow{}

\bibliography{biblio}

\end{document}




\section{Materials and Methods}
The experimental setup consists of a biaxial compressive test in plane strain conditions already described~\cite{lebouil.2014a}. The granular material is composed of glass beads of diameter $d = 70 - 110 \mu m$ and initial volume fraction $\approx 0.60$. The granular
material is enclosed between a latex membrane ($85 \times 55 \times  25~mm^3$) and a glass plate. A pump produces a partial vacuum inside the membrane, creating a confining
stress $-\sigma_3=30~kPa~$. The sample is placed in the biaxial apparatus where a compression is imposed in the vertical direction while the lateral side are maintained at a constant pressure. At the top, a moving plate exerts a compression of the sample at fixed velocity and the bottom plate is fixed. The velocity of the motor is of the order $v \sim \mu m.s^{-1}$ (different velocities are used, see main text for precise values), leading to a deformation rate $\dot{\epsilon_M} = v/L \sim 10^{-5} s^{-1}$ where $L=85~mm$. The origin of deformation $\epsilon_M=0$ is taken at the beginning of the contact between top plate and sample, but its precise value is without important because analysis considers only deformation increments.

The stress applied at the top of the sample is defined as $-\sigma_1 = -\sigma_3 + F/S$, where F
is the force measured by a sensor fixed to the upper plate, and
$S=55\times 25~mm^2$ is the section of the sample that we consider as constant during all the experiment.

Deformations are observed through the front glass plate using diffusing wave spectroscopy (DWS), a method already described elsewhere~\cite{erpelding2008,amon2017}. A laser beam at 532 nm is
expanded to illuminate the entire sample. The light undergoes
multiple scattering inside the granular material and we collect
the backscattered rays. The latter interfere and form a speckle
pattern. The image of the front side of the sample is recorded by
a $2352 \times 1728$ pixels camera. Speckle images are then subdivided
in square zones of size $16 \times 16$ pixels that we call metapixel.
The correlation between two images 1 and 2 is then calculated for each metapixel, and a map of correlation of $147 \times 108$ metapixels is obtained. The size of the metapixel correspond to $6.0~d \times 6.0~d$ on the sample.

\section{Spatiotemporal correlation function}
Average $\langle \cdot \rangle$ in equation~(3) is computed in the following way. The total length of the spatiotemporal series is of $7 \times 10^{4} \delta \varepsilon^*_M$. To smooth out the fluctuations of the strain rate at the level of the studied band, correlation functions are computed by slices of length $3 \times 10^{3} \delta \varepsilon^*_M$ in deformation interval and on position. Then the resulting functions are averaged on the whole duration of the series.

\section{Analysis of the Californian catalog}
\subsection*{Spatiotemporal dynamics}

\begin{figure}[h]
\centering
\includegraphics[width=0.4\columnwidth]{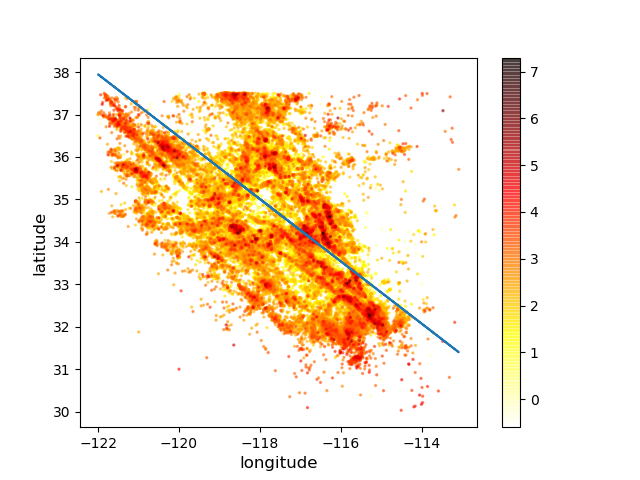}
\caption{Map of the events contained in the Californian catalog used for the analysis. The color of the dots represent the magnitude of the event (see colorbar). The blue line is obtained by a fitting procedure of all the data weighted by the magnitude of the events.}
\label{proj_catalog}
\end{figure}

Figure 2c  is obtained from a projection of the Earthquake Catalog for Southern California~\cite{url_catalog}. The events are represented as points in Figure~\ref{proj_catalog} with a color corresponding to their magnitude (see colorbar). To obtain a unidimensional time series, the position $x$ of the events along the general direction of the seismic activity (blue line in Fig.~\ref{proj_catalog}) are computed. Space is then binned in linear steps of $\Delta x=10$km and time in steps of $\Delta t= 10$ days. For each spatio-temporal window, the sum of the moments of the earthquakes occurring is computed:
\begin{equation*}
    M = \sum 10^{1.5(m_{w} + 6.07)},
\end{equation*}
giving the total moment released in the considered area and time interval. The values of those sums $M$ are represented with a logarithmic scale in Figure 2d (see colorbar on the right side of Figure 2d).

\subsection*{Correlation function} The average slip $\overline{D}$ during an earthquake is estimated using $M = \mu_0 S \overline{D}$ with $\mu_0$ the shear modulus and $S$ the fault area and the scaling relationship $M \propto S^{3/2}$~\cite{dearcangelis2016}.
\begin{equation*}
    \overline{D} \sim \frac{M}{M^{2/3}} \sim 10^\frac{m_w}{2}
\end{equation*}

The total slip values $\overline{D}$ are computed for spatiotemporal windows of sizes $\Delta x=1$km and $\Delta t= 1$ days. The correlation function $C(\delta x, \delta t)$ of those slip values is computed using an expression similar to Eq.~(3):
\begin{equation*}
C(\delta x,\delta t)=
\frac{
\langle \overline{D}(x,t)\times \overline{D}(x+\delta
x,t+\delta t) \rangle}
{\langle \overline{D}(x,t) \rangle \langle \overline{D}(x+\delta
  x,t+\delta t) \rangle} - 1
\end{equation*}

Averages are computed on both on space and time using the same procedure as the one used for experimental data.

\section{Deformation measurement with DWS}

Deformation are obtained using dynamic light scattering of a shear packing of spheres. Fig.1 shows a half-space filled of dielectric beads. The material is at rest excepted a zone where the strain is a pure shear.

\begin{figure}[h]
\centering
\includegraphics[width=0.4\columnwidth]{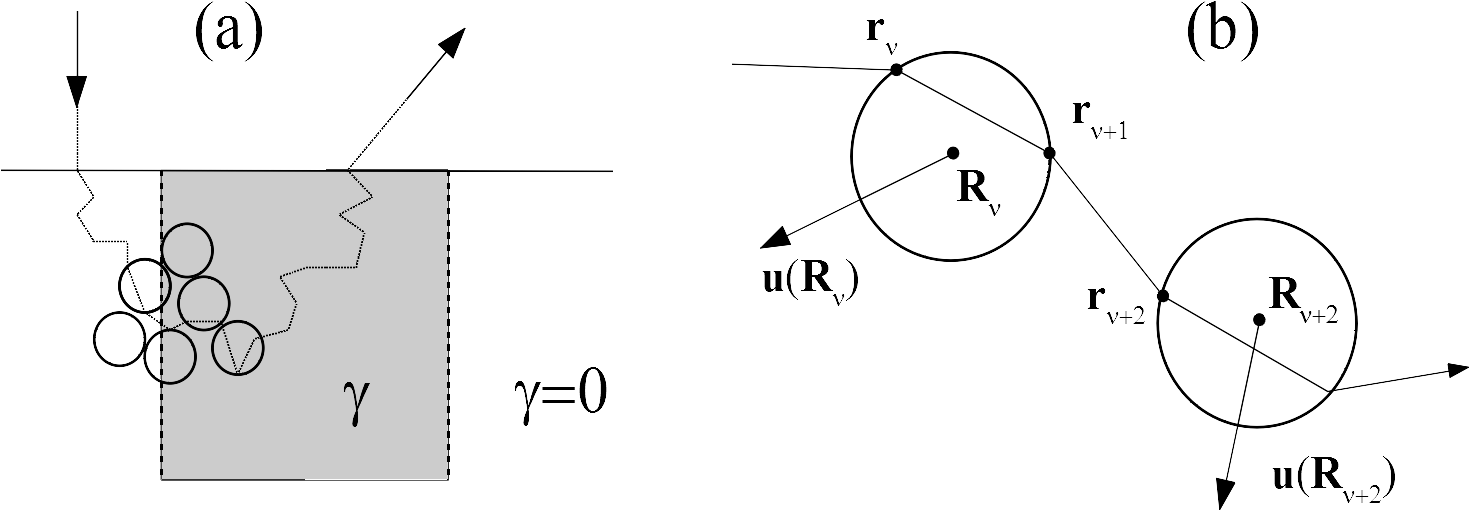}
\caption{(a) Model for inhomogeneous deformation. Only the material inside the gray zone is deformed. (b) A part of an optical path inside a glass bead assembly. }
\label{figDWS}
\end{figure}

\subsection*{Optical path}
A set of rays emerging at a given point of the surface is obtained using geometrical optics for the light propagation, with an algorithm explained elsewhere~\cite{crassous.2007}. The boundary conditions are vacuum for the outside material and an optical refractive index $n=1.51$ for the beads. The incident rays are perpendicular to the upper interface of the material.

A given path $i$ is decomposed into a succession of segments as shown on fig.\ref{figDWS}(b). Let's $\bfr_\nu$ the points at the interfaces between beads and the surrounding materials and $\bfk_\nu=2\pi n_\nu/\lambda_0$, where $n_\nu$ is the refractive index separating $\bfr_\nu$ and $\bfr_{\nu+1}$, and $\lambda_0$ the optical wavelength. The optical phase along the ray is :

\begin{equation}
\Phi^{(1)}_i=\sum_\nu \bfk_{\nu}\cdot\bigl[\bfr_{\nu+1}-\bfr_{\nu}\bigr]
\end{equation}

\subsection*{Deformation of optical path}

A optical ray is then deformed by moving beads. Let's $\bfR_\nu$ the center of the bead where the interface points $\bfr_\nu$ is located. Different points may belong to the same beads as shown on fig.\ref{figDWS}(b) where $\bfR_\nu=\bfR_{\nu+1}$. The center of every beads are then displaced of a quantity ${\bfu}(\bfR_{\nu})$, and the optical phase is :

\begin{equation}
\Phi^{(2)}_i=\sum_\nu \bfk_{\nu}\cdot\bigl[\bfr_{\nu+1}+{\bfu}(\bfR_{\nu+1})-(\bfr_{\nu}+{\bfu}(\bfR_{\nu})\bigr]
\end{equation}

The phase shift is then :
\begin{equation}
\Delta \Phi_i=\sum_\nu \bfk_{\nu}\cdot\bigl[{\bfu}(\bfR_{\nu+1})-{\bfu}(\bfR_{\nu})\bigr]
\end{equation}

For a segment of path that is sheared :

\begin{equation}
{\bfu}(\bfR_{\nu+1})-{\bfu}(\bfR_{\nu})=\doverline{\gamma}_\nu \cdot (\bfR_{\nu+1}-\bfR_{\nu})
\end{equation}
where $\doverline{\gamma}_\nu$ is the shear tensor of the segment.

Finally :
\begin{equation}
\Delta \Phi_i=\sum_\nu \bfk_{\nu}\cdot \bigl[\doverline{\gamma}_nu \cdot (\bfR_{\nu+1}-\bfR_{\nu})\bigr]
\end{equation}

\subsection*{Intensity correlation function}

The normalized correlation function of the scattering electric field is then computed as :

\begin{equation}
g_E=\frac
{\sum_i \exp^{j\Delta \Phi_i} \exp^{-s_i/l_a}}
{\sum_i \exp^{-s_i/l_a}}
\end{equation}

The factor $\exp^{-s_i/l_a}$, with $s_i$ the total optical length of the path $i$, and $l_a$ the absorption length, takes into account the absorption of light along the path.

Finally, the normalized intensity correlation function is :
\begin{equation}
g_I=\vert g_E \vert^2
\end{equation}

\subsection*{Homogeneous deformation}

\begin{figure}[h]
\centering
\includegraphics[width=0.4\columnwidth]{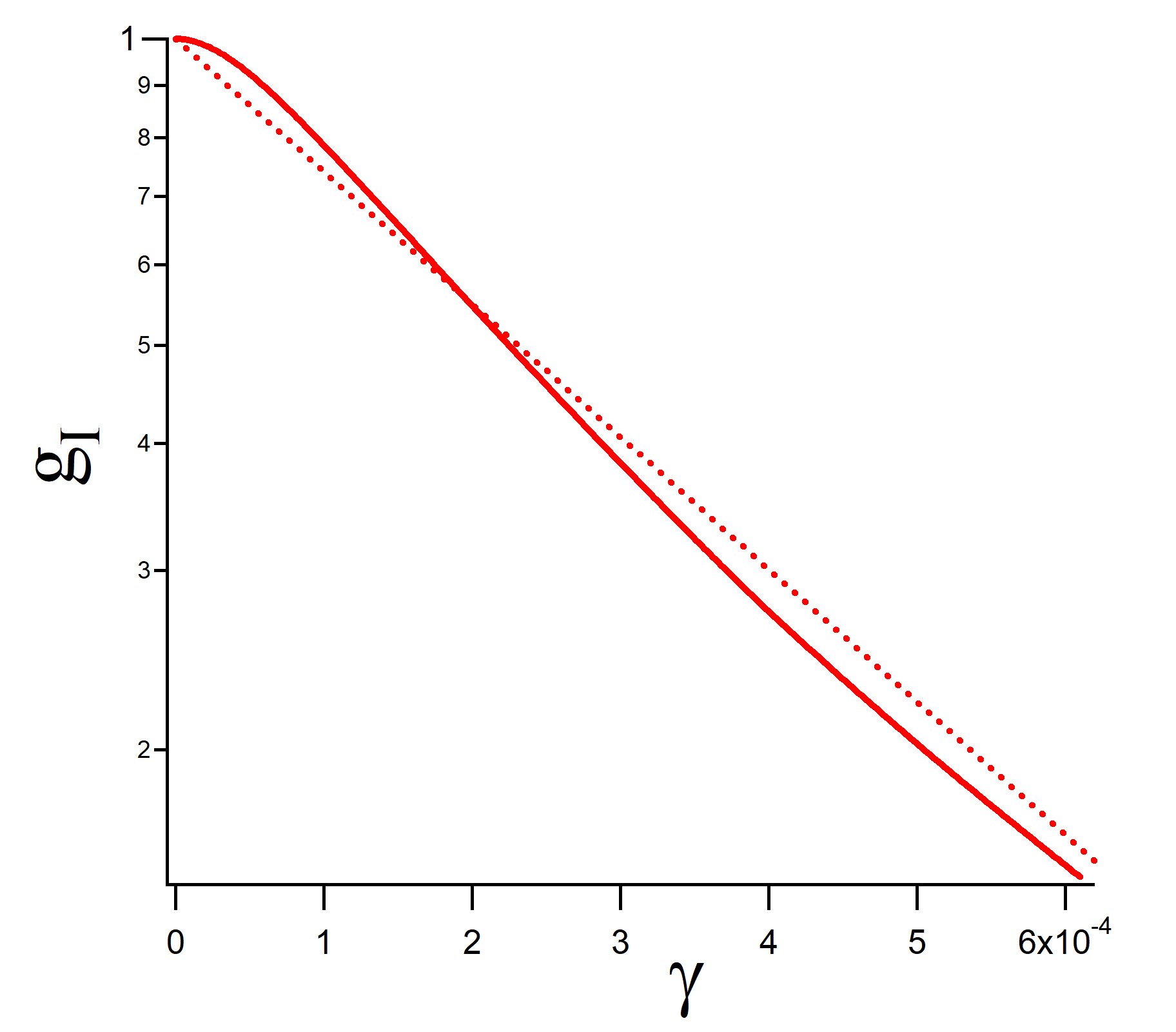}
\caption{Normalized intensity correlation function for an homogenous shear. Plain line is the result of numerical simulation, and the dotted line is $g_I=\exp{[-\gamma/\gamma_0]}$, with $\gamma_0=2.65\times10^{-4}$}
\label{fig_homo}
\end{figure}

For an homogeneous shear $(\doverline{\gamma}_\nu)_{ij}=\gamma \delta_{ix} \delta_{jy}$, we find $g_I \simeq \exp{[-\gamma/\gamma_0]}$, with $\gamma_0=2.65\times10^{-4}$. For the simulation, we used  the experimental values $d=90~\mu m$, $\lambda=0.532~\mu m$, and $l_a=44~mm$. The correlation decays exponentially with the amplitude of the strain it may be predicted with an analytical model~\cite{crassous.2007}. The "rounding" of the correlation function at small deformation observed in the lin-log plot of fig.\ref{fig_homo} is due to the presence of a finite absorption length.

\subsection*{Heterogeneous deformation}

\begin{figure}[h]
\centering
\includegraphics[width=\columnwidth]{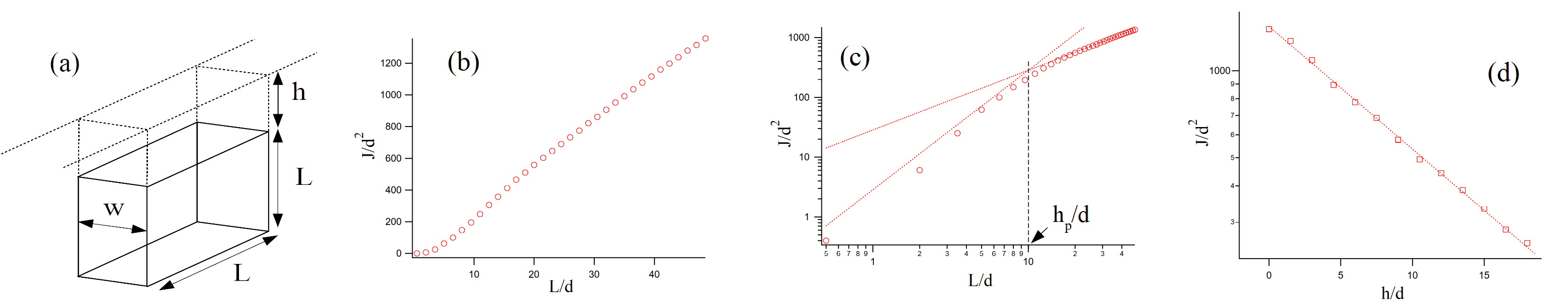}
\caption{(a)Definition of the sheared block. Only material inside a block of size $L\times L \times w$ situated to a depth $h$ with respect to surface is sheared. (b) $J/d^2$ as a function of $L$ ($w=15d$, $h=0$, $\gamma=4.7\times 10^{-4}$). (c) Same date as (b) in log-log plot. Dotted line are asymptotic behavior (see text for details). (d) $J/d^2$ as a function of $h$ ($w=15d$, $L=50d$, $\gamma=4.7\times 10^{-4}$). Dotted line is exponential decay  $J/d^2 \sim \exp{[-h/h_p]}$ with $h_p=10.2d$.}
\label{fig_hetero}
\end{figure}

We study the effect of heterogeneous deformation by considering blocks of sheared material embedded into non-deformed material. Figure \ref{fig_hetero}(a) shows a shear block of size $L\times L \times w$ at a depth $h$. The shear inside the block is $\gamma$. We perform ray tracing, and for every point of the surface we measure $g_I(\bfr)$, and we compute $J=\int_{\Sigma} -\ln[g_I(\bfr)] d\bfr$, where $\Sigma$ is a surface large compared to the defect. Fig.\ref{fig_hetero}(b-c) shows $J$ as a function of $L$ for a fixed width $w=15d$ and $h=0$. Results may be interpreted by remarking that: (i) $J$ is proportional to the {\it probed and deformed} volume $v$, (ii) a typical layer of depth $h_p\sim l^*$ is probed, (iii) for a homogeneous deformation $J/\int_S d\bfr =\gamma/\gamma_0$. It then follows that $J=(\gamma/\gamma_0) v/h_p$. If $L \gtrsim h_p$ and $h=0$, $v=L w h_p$, and then $J=(\gamma/\gamma_0) L w$. If $L \gtrsim h_p$ and $h=0$, then $v=L^2 w$, and $J=(\gamma/\gamma_0) L^2 w/h_p$. Those two limit behaviors are plotted as dotted lines on Fig.\ref{fig_hetero}(c). The cross-over between those limit behaviors occurs for $L=h_p$ which may be determined numerically as $h_p \simeq 10d$. The value of the finite depth of the probed layer may also be measured by computing $J$ for sheared zones situated below the surfaces. Fig.\ref{fig_hetero}(d) shows $J$ as a function of $h$. We measured that $J \sim \exp[-h/h_p]$ with $h_p\simeq 10d$.

\section{Threshold for definition of events.}

\begin{figure}[h]
\centering
\includegraphics[width=0.8\columnwidth]{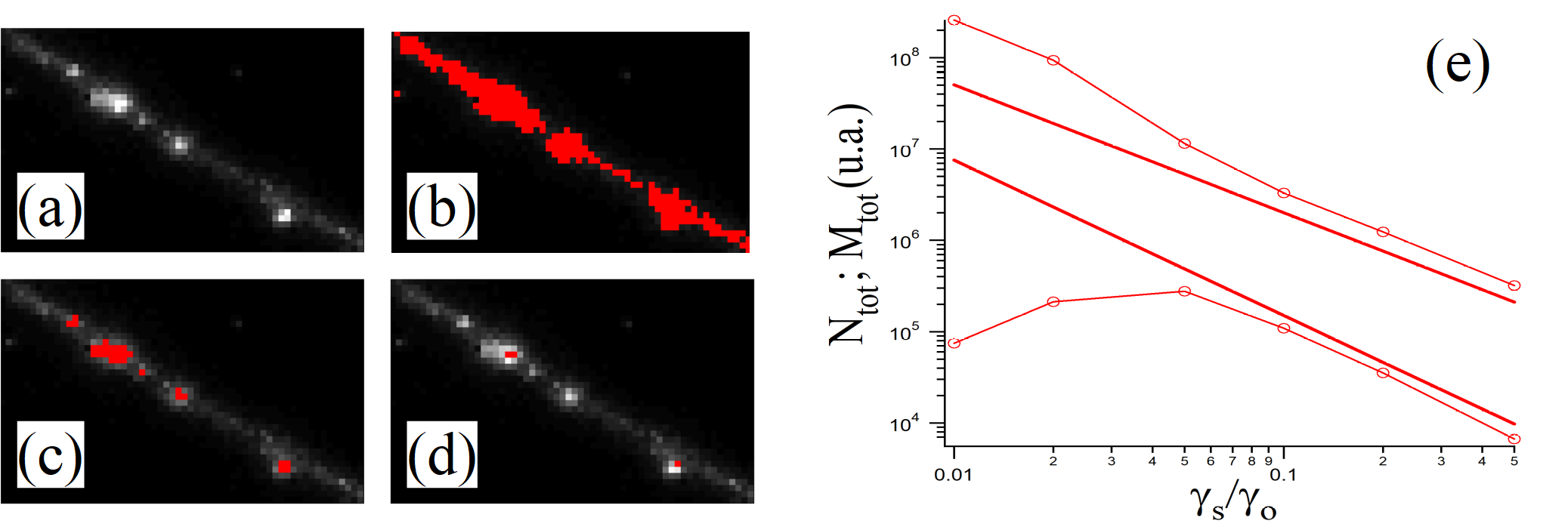}
\caption{(a) Image of the deformation. (b-c) Image with pixels above threshold $\gamma_s$ in red. (b):$\gamma_s=0.02 \gamma_0$; (c):$\gamma_s=0.1 \gamma_0$; (d):$\gamma_s=0.5 \gamma_0$. (e): Number of detected events $\circ$ and sum of the moment of events $\Box$ as a function of the threshold value. Lines are power laws for guide eyes.}
\label{fig_seuil}
\end{figure}

Events are defined as a set of connected of pixels with deformation over a given threshold. Starting for a intensity correlation map $g_I(\epsilon_M,\bfr)$, we compute microscopic deformation $\gamma_{m}(\bfr,\epsM)= - \gamma_0 \ln[g_I(\bfr,\epsM)]$ (see fig.\ref{fig_seuil}(a)). Pixels with deformation above a given threshold $\gamma_{m}(\bfr,\epsM)>\gamma_s$ are considered (fig.\ref{fig_seuil}(b-d)). A particle detection (function "Analyze Particles" of Fiji software \cite{Schindelin2012}) is then used, with a minimal size of $2$ pixels. A numbered list of cluster of pixels is then obtained. For choosing the threshold, we perform the analysis for different values of $\gamma_s$. We plotted on fig.\ref{fig_seuil}(e) the number of detected events. At low threshold the number of detected events decays because the different decorrelated zones collapse, whereas at high threshold the number of events decays because very few events are detected. The threshold is chosen to $\gamma_s=0.1 \gamma_0$ such as merging between events is low but number of events enough for statistics of moment and of aftershocks.

\section{Estimate of the ratio of dissipation R}

Let's $R=E_{ev}/E_{dis}$ the ratio between energy dissipated during events $E_{ev}$ over dissipated frictional energy $E_{dis}$. The energy dissipated by events of moment greater than $M$ is $E_{ev}(M)=\sum_{M_i\ge M} E_i$, with $E_i=\tau u_i L_i^2$. The detected events are located in a surface of shear band $L_b \times h_p$, with $h_p$ the depth of the shear band probed by DWS, and $L_b$ the length of the shear band under study. The total dissipated energy when the sliding of the two blocks is then $u_{tot}$ is then $E_{dis}=\tau L_b h_p u_{tot}$. Then $R=\sum_{M_i\ge M} u_i L_i^2 / L_b h_p u_{tot}$ where the sum is performed on detected events occurring on a length $L_b$ of shear band and during a total displacement $u_{tot}$ of the two blocks.

\section*{Movie SI1}
Left: deviatoric stress as a function of the macroscopic deformation. Right: the corresponding deformation map. The film shows different stages of the experiments, and data between them are not shown. In the manuscript, we analyze the deformation after yielding  $\varepsilon_M>6 \%$.

\bibliography{biblio,biblio_SM}